\documentclass[manuscript]{aastex}

\usepackage{graphicx}
\usepackage{amssymb}
\usepackage{amsmath}
\usepackage{epstopdf}
\newcommand{\kapand}{$\kappa$~And}
\newcommand{\kapandb}{$\kappa$~And~b}

\slugcomment{Accepted for publication in Astrophysical Journal Letters.}

\shorttitle{Super-Jupiter Orbiting \kapand}
\shortauthors{Carson et al.}

\begin{document}


\title{Direct Imaging Discovery of a `Super-Jupiter' Around the late B-Type Star 
  $\boldsymbol{\kappa}$~And\altaffilmark{\star}}


\author{J. Carson\altaffilmark{1,2},  C. Thalmann\altaffilmark{3,2},
  M. Janson\altaffilmark{4}, T. Kozakis\altaffilmark{1}, M. Bonnefoy\altaffilmark{2}, B. Biller\altaffilmark{2}, J. Schlieder\altaffilmark{2}, T. Currie\altaffilmark{5}, M. McElwain\altaffilmark{5}, 
    M. Goto\altaffilmark{6},
  T. Henning\altaffilmark{2}, W. Brandner\altaffilmark{2},
  M. Feldt\altaffilmark{2},  R. Kandori\altaffilmark{7},  M. Kuzuhara\altaffilmark{7,8}, L. Stevens\altaffilmark{1}, P. Wong\altaffilmark{1}, K. Gainey\altaffilmark{1}, M. Fukagawa\altaffilmark{9}, Y. Kuwada\altaffilmark{9},  T. Brandt\altaffilmark{4}, J. Kwon\altaffilmark{7}, L. Abe\altaffilmark{10}, S. Egner\altaffilmark{11}, C. Grady\altaffilmark{5}, O. Guyon\altaffilmark{11}, J. Hashimoto\altaffilmark{7}, Y. Hayano\altaffilmark{11}, M. Hayashi\altaffilmark{7}, S. Hayashi\altaffilmark{11}, K. Hodapp\altaffilmark{12}, M. Ishii\altaffilmark{11}, M. Iye\altaffilmark{7}, G. Knapp\altaffilmark{4}, T. Kudo\altaffilmark{11}, N. Kusakabe\altaffilmark{7}, T. Matsuo\altaffilmark{13}, S. Miyama\altaffilmark{14}, J. Morino\altaffilmark{7}, A. Moro-Martin\altaffilmark{15}, T. Nishimura\altaffilmark{11}, T. Pyo\altaffilmark{11}, E. Serabyn\altaffilmark{16}, H. Suto\altaffilmark{7}, R. Suzuki\altaffilmark{7}, M. Takami\altaffilmark{17}, N. Takato\altaffilmark{11}, H. Terada\altaffilmark{11}, D. Tomono\altaffilmark{11}, E. Turner\altaffilmark{4,18}, M. Watanabe\altaffilmark{19}, J. Wisniewski\altaffilmark{20}, T. Yamada\altaffilmark{21}, H. Takami\altaffilmark{7}, T. Usuda\altaffilmark{11}, M. Tamura\altaffilmark{7}}


\altaffiltext{$\star$}{Based on data collected at Subaru Telescope, which
  is operated by the National Astronomical Observatory of Japan.}
\altaffiltext{1}{Department of Physics \& Astronomy, College of Charleston, 58
  Coming  St., Charleston, SC 29424, USA}
\altaffiltext{2}{Max Planck Institute for Astronomy, Heidelberg, Germany}
\altaffiltext{3}{Astronomical Institute ``Anton Pannekoek'', University of Amsterdam,
  Science Park 904, 1098 XH Amsterdam, The Netherlands}
\altaffiltext{4}{Department of Astrophysical Sciences, Princeton University,
  NJ 08544, USA}
\altaffiltext{5}{ExoPlanets and Stellar Astrophysics Laboratory, Code 667,
  Goddard Space Flight Center, Greenbelt, MD 20771, USA}
\altaffiltext{6}{Universit\"ats-Sternwarte M\"unchen,
  Ludwig-Maximilians-Universit\"at, 81679 M\"unchen, Germany}
\altaffiltext{7}{National Astronomical Observatory of Japan, 2-21-1 Osawa, Mitaka, Tokyo 181-8588, Japan}
\altaffiltext{8}{Department of Earth and Planetary Science, The University of Tokyo, 7-3-1, Hongo, Bunkyo-ku, Tokyo 113-0033, Japan}
\altaffiltext{9}{Department of Earth and Space Science, Graduate School
of Science, Osaka University, 1-1 Machikaneyama, Toyonaka, Osaka
560-0043 Japan}
\altaffiltext{10}{Laboratoire Lagrange (UMR 7293), Universite de Nice-Sophia Antipolis, CNRS, Observatoire de la Cˆote d’Azur, 28 avenue Valrose, 06108 Nice Cedex 2, France}
\altaffiltext{11}{Subaru Telescope, 650 North A’ohoku Place, Hilo, HI 96720, USA}
\altaffiltext{12}{Institute for Astronomy, University of Hawaii, 640 N. A’ohoku Place, Hilo, HI 96720, USA}
\altaffiltext{13}{Department of Astronomy, Kyoto University, Kitashirakawa-Oiwake-cho, Sakyo-ku, Kyoto, Kyoto 606-8502, Japan}
\altaffiltext{14}{Hiroshima University, 1-3-2, Kagamiyama, Higashihiroshima, Hiroshima 739-8511, Japan}
\altaffiltext{15}{Departamento de Astrof\'isica, CAB (INTA-CSIC), Instituto
  Nacional T\'ecnica Aeroespacial, Torrej\'on de Ardoz, 28850, Madrid, Spain}
\altaffiltext{16}{Jet Propulsion Laboratory, California Institute of Technology, Pasadena, CA, 171-113, USA}
\altaffiltext{17}{Institute of Astronomy and Astrophysics, Academia Sinica, P.O. Box 23-141, Taipei 10617, Taiwan}
\altaffiltext{18}{Kavli Institute for Physics and Mathematics of the Universe, The University of Tokyo, 5-1-5, Kashiwanoha, Kashiwa, Chiba 277-8568, Japan}
\altaffiltext{19}{Department of Cosmosciences, Hokkaido University, Kita-ku, Sapporo, Hokkaido 060-0810, Japan}
\altaffiltext{20}{HL Dodge Department of Physics \& Astronomy, University of Oklahoma, 440 W Brooks St, Norman, OK 73019 USA}
\altaffiltext{21}{Astronomical Institute, Tohoku University, Aoba-ku, Sendai, Miyagi 980-8578, Japan}


\begin{abstract}
We present the direct imaging discovery of an extrasolar planet, or possible low-mass brown dwarf, at a projected separation of 55 $\pm$ 2 AU ($1\farcs058 \pm 0\farcs007$) from the B9-type star \kapand.  The planet was detected with Subaru/HiCIAO during the SEEDS survey, and confirmed as a bound companion via common proper motion measurements.  Observed near-infrared magnitudes of $J$ = 16.3 $\pm$ 0.3, $H$ = 15.2 $\pm$ 0.2, $K_\mathrm{s}$ = 14.6 $\pm$ 0.4, and $L^\prime$ = 13.12 $\pm$ 0.09 indicate a temperature of $\sim$1700 K.   The galactic kinematics of the host star are consistent with membership in the Columba association, implying a corresponding age of 30$^{+20}_{-10}$ Myr.  The system age, combined with the companion photometry, points to a model-dependent companion mass $\sim$12.8 $M_\mathrm{Jup}$.   
The host star's estimated mass of 2.4--2.5 $M_{\rm \odot}$ places it among the most massive stars ever known to harbor an extrasolar planet or low-mass brown dwarf.  While the mass of the companion is close to the deuterium burning limit, its mass ratio, orbital separation, and likely planet-like formation scenario imply that it may be
best defined as a `Super-Jupiter' with properties similar to other
recently discovered companions to massive stars.


\end{abstract}





\section{Introduction}

Stellar mass is emerging as one of the most important parameters in determining the properties of planetary systems, along with stellar metallicity. Radial velocity surveys have indicated that the frequency of giant planets increases with the mass of the stellar host \citep{john10}, and many of the roughly dozen exoplanets that have been directly imaged so far have had A-type stellar hosts \citep[e.g.,][]{maro08,lagr09}, despite such large stars being in the small minority of surveyed targets. These results have motivated targeted imaging surveys for planets around massive stars \citep[e.g.,][]{jans11b}. The increase in planet frequency with host star mass can be readily explained theoretically, through the consideration that more massive stars are likely to have more massive disks \citep{mord12}. On the other hand, massive stars also feature an increased intensity of high-energy radiation, which may significantly shorten the disk lifetime due to photoevaporation, and thus decrease the time window in which giant planets are allowed to form. This raises the question whether there is a maximum stellar mass above which giant planets are unable to form. 

In this Letter, we report the discovery of a $\sim$12.8~$M_{\rm Jup}$ companion to the $\sim$2.5~$M_{\rm \odot}$ star \kapand{}, the most massive star to host a directly detected companion below or near the planetary mass limit. In the following, we describe the acquisition, reduction and analysis of the data used for detection, confirmation, and basic characterization of the companion, \kapandb{}.

\section{Observations and Data Reduction}

\begin{figure}
\centering
\includegraphics[scale=0.21]{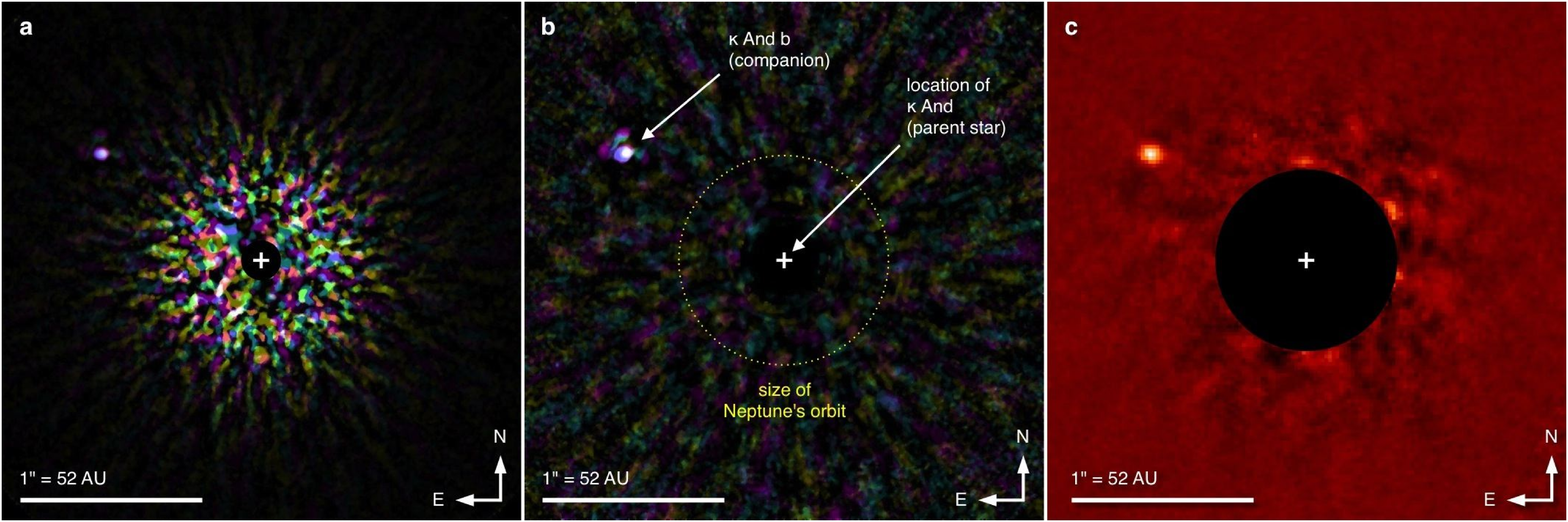}      
\caption{\textit{Left}: \textit{JHK} false-color image of \kapandb{} after LOCI/ADI data reduction, for the 2012 July observations.   \textit{Center}: A corresponding signal-to-noise map created from the left frame. The S/N ratio is calculated in concentric annuli around the star. The white plus sign in each panel marks the location
  of the host star \kapand; the black disks designate the regions
  where field rotation is insufficient for ADI.  White features indicate where the signal is roughly equally strong in all wavelengths; colored features indicate where the signal is mismatched between wavelengths, and is often indicative of residual noise.  The lobes around \kapandb{} result from the Airy pattern produced by the Subaru AO188 system.  \textit{Right}: $L'$-band image of \kapandb{} from the 2012 July observations.
\label{fig1}}
\end{figure}

\begin{figure}
\centering                                                
\includegraphics[scale=1.1]{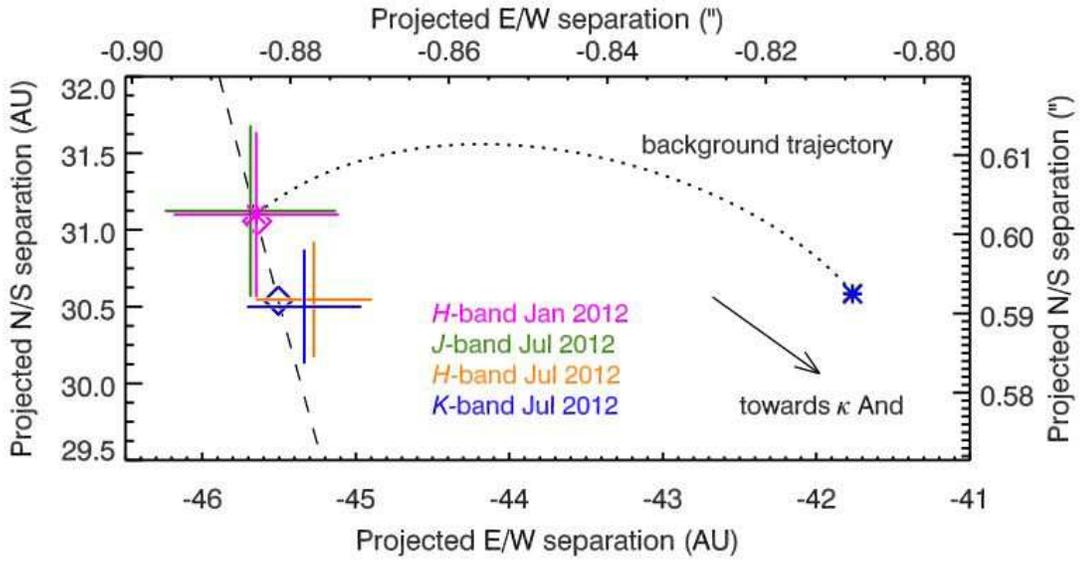}
\caption{Proper-motion analysis of \kapandb.  The dotted curve designates the predicted parallactic and proper motion between epochs, if the detected January source were a background star.  The dashed line indicates an example bound, orbital path of \kapandb{} consistent with the observational data.  The diamond symbols represent the predicted January and July astrometric measurements for \kapandb{}, if it follows the dashed orbital path.   \kapandb{} is clearly inconsistent with background behavior and instead demonstrates common proper motion with the host star. \label{fig2}} 
\end{figure}

Observations of the \kapand{} system extended over a period of seven months (January - July 2012) and were carried out on Subaru Telescope.  $JHK$ images were collected with AO188 \citep{haya10} coupled with HiCIAO \citep{hoda08}.      $L^\prime$ measurements were carried out with AO188 coupled with the Infrared Camera and Spectrograph \citep[IRCS;][]{toku98}.   Figure~\ref{fig1} displays the multi-wavelength images of the newly discovered companion. 
Table~\ref{tbl-1} provides a summary of the experimental measurements, as well as relevant values from the literature.  Figure~\ref{fig2} shows observed astrometric positions of \kapandb{} as compared with expected motion of an unrelated background star.  The sub-sections below describe the observations in greater detail.

\subsection{Subaru HiCIAO/AO188 $JHK$ Imaging}

We first detected \kapandb{} using AO188 coupled with HiCIAO
on Subaru Telescope on January 1, 2012, as part of the SEEDS survey \citep{tamu09}.
The observations used a 20$^{\prime\prime}$ $\times$ 20$^{\prime\prime}$ field of view, 9.5\,mas pixels, and an opaque 0\farcs6-diameter coronagraphic mask, which helped keep the saturation radius $<$ $0\farcs5$. The images were taken 
in the near infrared ($H$-band, 1.6\,$\mu$m),
where young substellar objects are expected to be bright with thermal
radiation \citep{bara03}.
Pupil tracking was used to enable angular differential imaging \citep[ADI;][]{maro06}.

Data reduction of the 46 exposures of 5\,s revealed, at $23\,\sigma$ 
confidence, a
pointlike source at 1\farcs07 separation.  Follow-up observations in $J$ (1.3\,$\mu$m; 177 exposures of 10\,s), $H$ (1.6\,$\mu$m; 171 exposures of 8\,s), and $K_\mathrm{s}$ (2.2\,$\mu$m; 135 exposures of 10\,s), collected on July 8--9, 2012, using the same observing setup, re-detected the source at $6$, $28$, and $49\,\sigma$ confidence levels, respectively.  Unsaturated images of the primary, taken immediately before and after each filter's observing sequence, and using a neutral density filter (0.866\% for $H$, 1.113\% for $K_\mathrm{s}$, and 0.590\% for $J$) provided photometric calibration. 

To optimize the ADI technique, we first reduced the data using a locally optimized combination of images algorithm \citep[LOCI;][]{lafr07}.  HiCIAO observations of M5, combined with distortion-corrected images obtained with the Advanced Camera for Surveys (ACS) on the Hubble Space Telescope, enabled accurate pixel scale calibration to within 0.2\%; the ACS astrometric calibration was based on \citet{vand07}.     Figure~\ref{fig1} (left and middle) presents a \textit{JHK} false-color image and corresponding signal-to-noise (S/N) map after the ADI/LOCI data reduction.

Given the relatively high S/N ratios and the known difficulties in quantifying the
impact of LOCI on planet photometry and astrometry, we also performed a
classical ADI reduction \citep{maro06} with \texttt{mean}-based point-spread function
(PSF) estimation and frame co-adding.  Unsharp masking on the spatial
scale of 35 pixels ($\approx$ 7\,FWHM) was applied to the final image
to flatten the residual background.   The planet signal was
recovered with S/N ratios comparable (within 10\%) to the LOCI reduction for all the July data
sets.  For the somewhat lower quality January data, the measured S/N reduced from about $23\,\sigma$ to $7\,\sigma$.  

To achieve unbiased photometry and astrometry, we extracted the combined \kapand{} PSF (S/N $>$ 1000) from the neutral density images, and placed it on an empty image frame at the location of \kapandb{}.  Applying the same unsharp masking and ADI reduction to this data as we did for the science data, we simulated the parallactic angle evolution, as recorded in the science frames.  The resulting processed PSF acted as the photometric and astrometric reference for \kapandb{}. The only non-linear step in this process was the \texttt{median}-based unsharp masking, but the large spatial scale ($\approx$ 7\,FWHM) ensured that subtraction effects were minimal.  

    We calibrated the astrometry by cross-correlating the \kapandb{} signal with the processed calibration PSF.  We estimated the uncertainty in the \kapandb{} center to be $\mathit{FWHM}/(S/N)$, following \citet{came08}.   The uncertainties in the final relative astrometry were dominated by our ability to determine the host star center, which was carried out through Moffat fitting of each individual exposure.  We conservatively estimated the uncertainty of the Moffat fit at 0.75 pixels (7 mas).  For confirmation, we applied Moffat fitting and peak fitting to unsaturated data of \kapand{} and found that the methods agreed at the 0.5\,$\sigma$ level.  The photometric uncertainties were calculated as a combination of (1) representative noise in an annulus, centered on the host star, with a radius equal to the companion, (2) photometric variability in the neutral density calibration images, which yielded effective accuracies of 7--11\% for the combined datasets, and (3) uncertainties in the \textit{JHK} magnitudes of \kapand.  


\subsection{Subaru IRCS/AO188 $L^\prime$ Imaging}

On July 28, 2012, we followed the \textit{JHK} observations with $L^\prime$-band observations (3.8\,$\mu$m; 50 exposures of 30\,s)  using AO188 coupled with the Infrared Camera and Spectrograph on Subaru Telescope.  We employed a 10\farcs5 $\times$ 10\farcs5  field of view, 20.6 mas pixel scale and no coronagraph. The host star saturated out to $\sim$ $0\farcs1$.  The dithered observations, carried out in ADI mode, were divided into two identical sequences bracketing observations of the star HR 8799, which provided the photometric calibration \citep{maro08}. Observations of a third star, S810-A, were collected before the science observations as a secondary calibration check \citep{legg03}. 

We sky-subtracted each image using a median combination of frames taken at the other dither positions.  To help maximize the high-contrast sensitivity, we processed the data using an ``adaptive'' LOCI process (A-LOCI; \citealt{curr12}).  We also employed a moving pixel mask, where the LOCI algorithm is prevented from using, in PSF construction, 
pixels lying within the subtraction zone (see \citealt{lafr07} for details).  Figure~\ref{fig1} (right) shows the final image.
  
To quantify the  \kapandb{} throughput, we 
used fake point sources added to the image and processed with the same algorithm settings.  As an additional check on our flux calibration, 
we determined the relative brightness between the HR 8799 bcd planets (all detected at S/N $>$ 7--10) using identical procedures, and confirmed its agreement with published values \citep{curr11}.  The independent calibrations all yielded 
self-consistent results, ensuring confidence in the 22\,$\sigma$ detection of \kapandb{} in $L^\prime$.  As a final check, we re-processed the $L^\prime$-band data using a more classical ADI method, similar to that described for the $JHK$ data set, and achieved consistent results.    While the July $L^\prime$ astrometry was consistent with the July \textit{JHK} results, we refrained from including it in our proper motion analysis, due to our possession of poorer-quality astrometric calibration.




\section{Host Star Properties}
\kapand{} is a B9 IV star \citep{wu11} located at a distance of 52.0 pc \citep{perr97}.  \citet{fitz05} report a temperature of 11,400 $\pm$ 100~K with a sub-solar metallicity of $[$Fe/H$]$ = $-$0.36 $\pm$ 0.09, while independent measurements by \citet{wu11} report values of 10,700 $\pm$ 300~K and $-$0.32 $\pm$ 0.15.  Given the star's spectral classification, the measured low metallicity is likely due to the details of the star's accretion and atmospheric physics, as opposed to a true, initial, low metallicity \citep{gray02}. We estimate a mass of 2.4--2.5 $M_{\rm \odot}$ using the published temperature and evolutionary tracks from \citet{ekstr12}.    
Table~\ref{tbl-1} summarizes the host star properties.  

\citet{zuck11} proposed \kapand{} to be a member of the $\sim$30 Myr old Columba association.  To further investigate \kapand's likely membership in Columba we: (1) independently calculated its Galactic kinematics from astrometry available in the literature \citep{perr97, zuck11} and compared these to the young local associations reported in \citet{torr08}, and (2) calculated its membership probability in these associations using the Bayesian methods of  \citet{malo12}.  Our analyses showed that the star's kinematics imply a $>$95\% probability of the star being part of the Columba association.      

As an additional check, we compared the \kapand $B-V$ color and absolute $V$
magnitude \citep{perr97, vanl09} with members of clusters and associations with ages ranging
from $\sim$15--700 Myr.  These include Lower Centaurus Crux, $\alpha$ Per, Pleiades, Coma
Ber, Hyades, Praesepe, and young local associations \citep{torr08, vanl09}.  The
color--magnitude analysis showed that \kapand{} is consistent with other early-type
stars having ages $\sim$20--120 Myr. The results of our analyses are consistent with the conclusions
reported in \citet{zuck11};  \kapand's age range and kinematics suggest it is a member in the Columba association.  We therefore adopt a system age of 30$^{+20}_{-10}$ Myr (following \citealt{maro10}) for all
subsequent analyses.    

\section{Results}

\subsection{Proper Motion Analysis}
Located 52.0 pc from the Sun, \kapand{} exhibits proper motion of 83.5 mas/yr \citep{perr97}, enabling an effective test to distinguish bound companions from unrelated background stars.   The \kapand{} proper and parallactic motion translate to 76 mas ($\sim$8 HiCIAO pixels) of net movement over the 6 month period between epochs.  As shown in Figure~\ref{fig2}, the companion exhibits common proper motion with the host star, and deviates from expected background star motion by $7\,\sigma$.  In addition to this $7\,\sigma$ deviation in the magnitude of motion, the observed direction of motion and scatter in astrometry are completely inconsistent with that of a background star.        

\subsection{Physical Properties of \kapandb}
\begin{figure}
\centering
\includegraphics[scale=0.7]{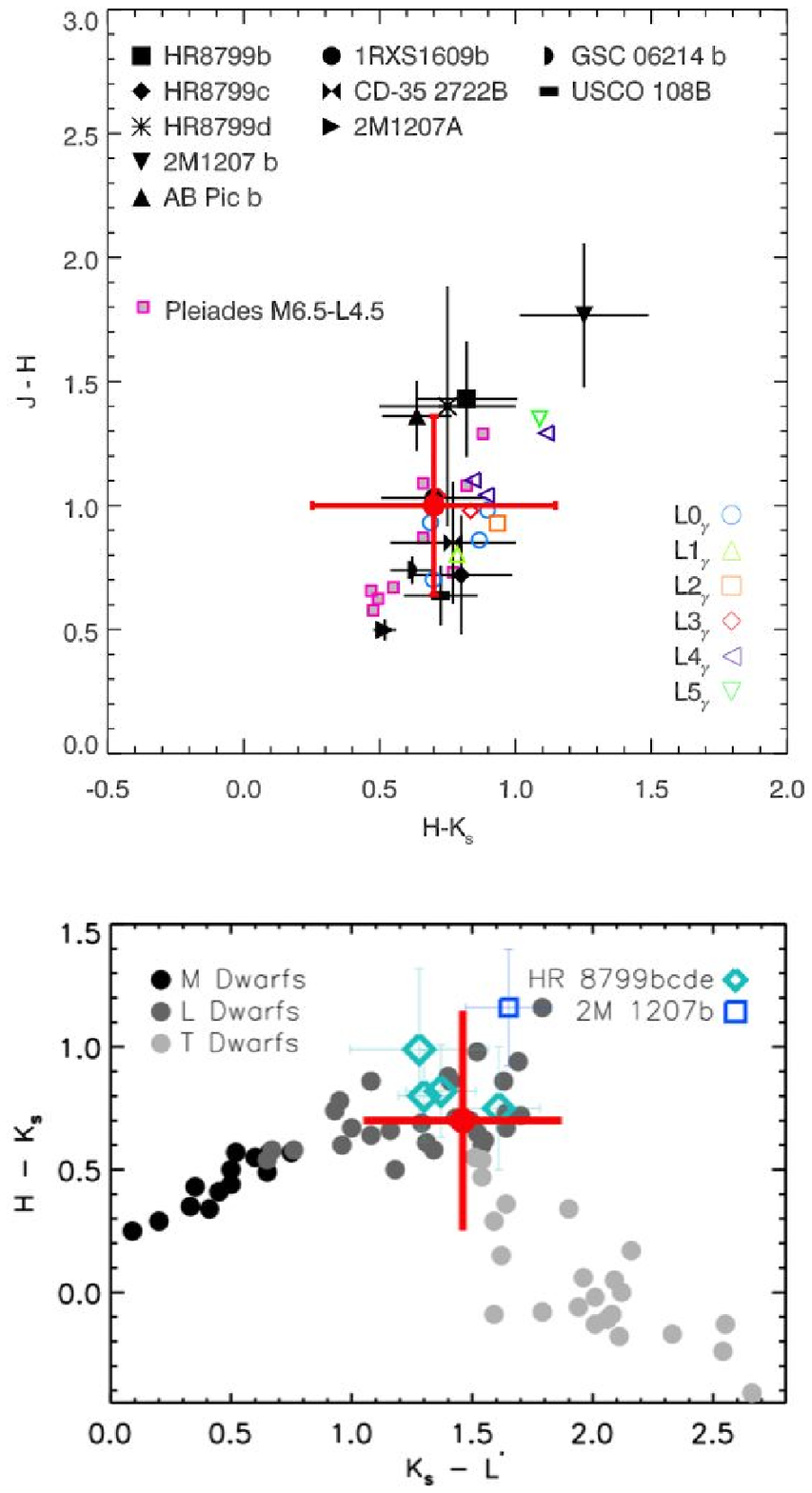}                                                
\caption{Position of \kapandb{} colors (red points) with respect to reference objects.  Top plot includes benchmark substellar companions: HR~8799~bcd \citep{maro08}, 2M1207~Ab \citep{chau04}, AB~Pic~b \citep{chau05}, 1RXS1609~b \citep{lafr08}, CD-35~2722~B \citep{wahh11}, GSC~06214~b \citep{irel11}, and USCO~108~AB  \citep{beja08}.  It also contains L dwarfs with spectral features indicative of reduced surface gravity (\citealt{cruz09}; \citealt{fahe12}), and Pleiades M-L dwarfs \citep{biha10}.   Bottom plot includes M, L, and T field dwarfs \citep{legg02}, HR~8799~bcde (\citealt{curr11}; \citealt{skem12}), and 2M1207~b \citep{chau04}.   \label{fig3}}
\end{figure}

\begin{figure}
\centering
\includegraphics[scale=0.6]{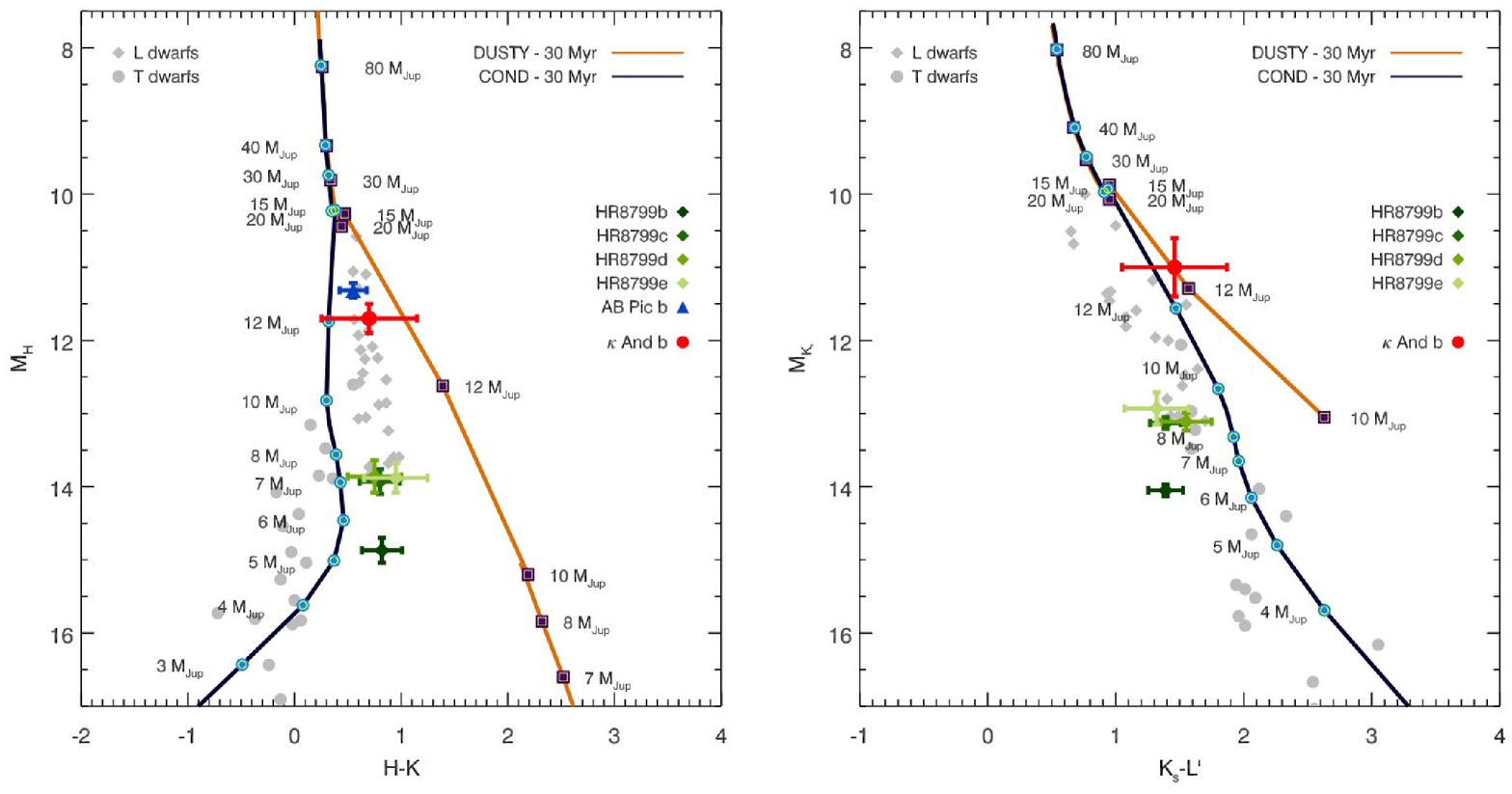}                                                 
\caption{\kapandb{} colors and absolute magnitudes (red points) compared with \texttt{DUSTY} and \texttt{COND} evolutionary tracks \citep{bara03, chab00}.      \label{fig4}}
\end{figure}

Figure~\ref{fig3} shows that the \kapandb{} colors are most consistent with cloudy L dwarfs and overlap with several other benchmark exoplanets and low-mass companions, including HR~8799~bcd, AB~Pic~b, and 1RXS1609~b.  
 Figure~\ref{fig4} compares \kapandb{} colors and absolute magnitudes with \texttt{DUSTY} and \texttt{COND} evolutionary tracks \citep{bara03, chab00}, as well as low-mass companions around HR 8799 and AB Pic.  The plots show \kapandb{} as well situated between HR~8799~cde and AB~Pic~b.  Its infrared colors are slightly bluer than those of typical field L dwarfs, possibly indicating a low surface gravity \citep{cruz09}.  However, improved photometry is required to confirm whether this color deviation is real. 

The estimated temperature of \kapandb{} suggests that its atmospheric properties should align more closely with those of the 
{DUSTY} models (see discussions in \citealt{chab00}).
In deriving a mass estimate from this track, we rely on the July $H$-band magnitude because (1) alternative $J$-band and January $H$-band measurements have higher uncertainties, (2) $K_\mathrm{s}$-band mass estimates are more sensitive to atmospheric composition (see e.g.\ \citealt{jans11}), and (3) $L^\prime$-band mass estimates have been less thoroughly tested with experimental data, and are more sensitive to age uncertainties for this age and magnitude range (see \citealt{chab00}).

Based on the July $H$-band magnitude of 15.2 $\pm$ 0.2, the estimated age of 30$^{+20}_{-10}$ Myr, a parallax of 19.2 $\pm$ 0.7 \citep{perr97}, and the \texttt{DUSTY} evolutionary models, we calculate a mass of 12.8$^{+2.0}_{-1.0}$ $M_{\rm Jup}$ and a temperature of 1680$^{+30}_{-20}$\,K.  As a consistency check, we calculate the predicted $J$$K_\mathrm{s}$$L^\prime$ magnitudes based on the estimated 1680~K temperature, the 20--50 Myr system age, and the \texttt{DUSTY} evolutionary models. This yields $J$ = 16.5--16.8, $K_\mathrm{s}$ = 14.2--14.4, and $L^\prime$ = 13.1--13.2 mags, all of which are in agreement with our measured multiband photometry.  Additionally, the two epochs of $H$-band photometry are in agreement with one another. 
Table~\ref{tbl-1} summarizes the complete properties of \kapandb{}.  

While the {DUSTY} models are likely the more relevant, we estimate a possible alternative mass using the {COND} evolutionary tracks.  In this scenario, we determine a mass of 11.5$^{+2.4}_{-1.2}$ $M_{\rm Jup}$ and a temperature of 1640$^{+40}_{-20}$\,K.  More recent evolutionary models by \citet{spie12} offer alternative ``Warm Start'' scenarios that consider formation with lower levels of initial entropy.  While these models do not consider combinations of mass and temperature similar to that of \kapandb{}, they do predict generally higher masses than that of the {DUSTY} and {COND} models.  In the case of \kapandb{}, such models place the most probable mass at a value above the typical deuterium burning limit.  While we currently adopt a nominal mass estimate of 12.8$^{+2.0}_{-1.0}$ $M_{\rm Jup}$ for the analyses in this discovery paper (based on the {DUSTY} models), we defer a deeper investigation of companion mass for a follow-up paper, where we will focus on a more thorough comparison of multiband photometry with synthetic spectra.

\subsection{Orbital Properties of \kapandb}
We estimate the semimajor axis of \kapandb{} from its observed separation.
Assuming a uniform eccentricity distribution of 0 $<$ \textit{e} $<$ 1, and random viewing angles, \citet{dupu10} compute a median correction factor between projected separation and semimajor axis
of 1.1$^{+0.91}_{-0.36}$.  Using this relation, we derive a semimajor axis of 61$^{+50}_{-20}$ AU based on its projected separation of 55.2 AU (1\farcs07) in January 2012.  

\subsection{Possible Secondary Companions}
The $H$-band sensitivity levels (see Section 2.1) allow us to rule out secondary companions with temperatures similar or warmer than that of \kapandb{}, for separations greater than 0\farcs9 (46 projected AU).  For the \kapandb{} separation (1\farcs1) and beyond, we may rule out secondary companions with masses $\geq$ 11.7 $M_\mathrm{Jup}$, assuming a 30 Myr system age and the \texttt{DUSTY} evolutionary models.

\section{Discussion}
$\kappa$~And is the most massive star to host a directly imaged planet, or brown dwarf near the deuterium burning boundary.
The mass ratio between $\kappa$~And~b and its host is $\sim$0.5\%, similar to the $\sim$0.4\% ratios of the $\beta$~Pic and HR~8799 planets \citep{lagr09,maro08}. In comparison, this value is noticeably smaller than those of reported directly imaged planets around 1RXS~1609 \citep{lafr08} and 2M~1207 \citep{chau04}. The projected separation of $\kappa$~And~b is also intermediate between the two outer planets in HR~8799. The similarities  between $\kappa$~And~b, $\beta$~Pic, and HR~8799 could imply a similar formation mechanism, which may be distinct from recently discovered brown dwarf companions of approximately an order of magnitude larger mass ratios \citep[e.g., GJ~758~B;][]{thal09} or semimajor axes \citep[e.g., HIP~78530~B;][]{lafr11}. Strengthening the possibility of a planet-like formation for Kap And b, theoretical models \citep[e.g.][]{rafi11} show that, for a minimum mass solar nebula, the region of the primordial disk where core accretion formation of giant planets can occur overlaps with the separation range of $\kappa$~And~b.  Furthermore, this formation mechanism may be significantly enhanced for a star as massive as $\kappa$~And, assuming it had a correspondingly more massive protoplanetary disk.  
 Further studies will be needed to more stringently constrain the population properties of planets and brown dwarfs on intermediate and wide orbits.

The best-fit mass of $\kappa$~And~b lies just below the deuterium burning limit according to conventional evolutionary models, but may be above this limit if initial entropy at formation is lower than such models assume \citep{spie12}. This leads to an ambiguity in whether the companion can be classified as an ``exoplanet'' by the present IAU definition. Such a classification scheme can however be misleading, given that $\kappa$~And~b may well have formed in the same way as previously imaged planets, regardless of whether its mass falls just below or above this limit. Indeed, radial velocity studies have shown that massive stars tend to have massive planets, sometimes with companions having masses above the deuterium burning limit \citep[e.g.][]{lovi07} and which apparently form a high-mass tail of a lower-mass planetary population (e.g.~\citealt{hekk08}). On the other hand, formation history can be difficult to assess in individual cases. In order to avoid these uncertainties, we simply classify $\kappa$~And~b as a `Super-Jupiter', which we take to mean a group of objects that includes the previously imaged planets around  HR~8799 and $\beta$~Pic as well as the most massive radial velocity planets, and which one might suspect have formed in a similar way to lower-mass exoplanets, but for which this has not necessarily been unambiguously demonstrated. This suggested class includes substellar objects with masses at or moderately above the deuterium burning limit, but excludes objects with orbital separations well beyond a typical disk truncation radius, or systems with mass ratios more indicative of a binary-like formation.

\acknowledgments

The authors thank David Lafreni{\`e}re for providing the source code for his LOCI algorithm, the anonymous referee for useful comments, and Subaru Telescope staff for their assistance.  The authors thank David Barrado and the Calar Alto Observatory staff for their efforts at carrying out supplementary observations of the host star. 
This work is partly supported by a Grant-in-Aid for Science Research in a Priority Area from MEXT, Japan, and the U.S. National Science Foundation under Award No. 1009203 (J.C., T.K., P.W., K.G.), 1008440 (C.G.), and 1009314 (J.W.).  The authors recognize and acknowledge the significant cultural role and reverence that the summit of Mauna Kea has always had within the indigenous Hawaiian community.  We are most fortunate to have the opportunity to conduct observations from this mountain.



{\it Facilities:} \facility{Subaru (HiCIAO, IRCS, AO188)}.

\clearpage



\clearpage

\clearpage

\clearpage

\begin{deluxetable}{l c c}
\tabletypesize{\scriptsize}
\tablecaption{Properties of the \kapand{} System \label{tbl-1}}
\tablewidth{0pt}
\tablehead{
\colhead{Property}  & \colhead{Primary} & \colhead{Companion} \\ 
}
\startdata
Mass  &  2.4--2.5\,$M_{\rm \odot}$\tablenotemark{a} & 12.8$^{+2.0}_{-1.0}$ $M_{\rm Jup}$\tablenotemark{b}  \\ 
$T_\mathrm{eff}$ &  11,400 $\pm$ 100 K\tablenotemark{c} &  1,680$^{+30}_{-20}$ K\tablenotemark{b}  \\
 &  10,700 $\pm$ 300 K\tablenotemark{d} &  ---  \\
Spectral Type  &  B9 IV\tablenotemark{d} &  L2--L8\tablenotemark{e}  \\ 
Age (Myr)  &  30$^{+20}_{-10}$\tablenotemark{f} &  ---  \\ 
Parallax (mas)  &  19.2 $\pm$ 0.7\tablenotemark{g} &  ---  \\ 
Fe/H  &  $-0.36$ $\pm$ 0.09\tablenotemark{c} &  ---  \\ 
&  $-0.32$ $\pm$ 0.15\tablenotemark{d} &  ---  \\ 
$\log g$  &  4.10 $\pm$ 0.03\tablenotemark{c} &  ---  \\ 
  &  3.87 $\pm$ 0.13\tablenotemark{d} &  ---  \\ 
$J$ (mag)  &  4.6 $\pm$ 0.3\tablenotemark{h} &  16.3 $\pm$ 0.3  \\
$H$ (mag)  &  4.6 $\pm$ 0.2\tablenotemark{h} &  15.2 $\pm$ 0.2  \\
$K_\mathrm{s}$ (mag)  &  4.6 $\pm$ 0.4\tablenotemark{h} &  14.6 $\pm$ 0.4  \\
$L^\prime$ (mag)  &  --- &  13.12 $\pm$ 0.09  \\ 
$\Delta$$J$ (mag)  &  --- &  11.6 $\pm$ 0.2  \\ 
$\Delta$$H$ (mag) &  --- &  10.64 $\pm$ 0.12  \\ 
$\Delta$$K_\mathrm{s}$ (mag)  & --- &  10.0 $\pm$ 0.08  \\ 
$M_{J}$ (mag)  &  1.0 $\pm$ 0.3\tablenotemark{i} &  12.7 $\pm$ 0.3  \\ 
$M_{H}$ (mag)  &  1.0 $\pm$ 0.2\tablenotemark{i} &  11.7 $\pm$ 0.2  \\ 
$M_{K_\mathrm{s}}$ (mag)  &  1.0 $\pm$ 0.4\tablenotemark{i} &  11.0 $\pm$ 0.4  \\ 
$M_{L^\prime}$ (mag)  &  --- &  9.54 $\pm$ 0.09  \\ 
Astrometry on 1 January 2012 ($H$-band):  \\
--- Proj.\ sep. ($^{\prime\prime}$) & --- & 1.070 $\pm$ 0.010 \\
--- Proj.\ sep. (AU) & --- & 56 $\pm$ 2\tablenotemark{j} \\
--- Position angle ($^\circ$) & --- & 55.7 $\pm$ 0.6 \\
Astrometry on 8 July 2012 ($H$-band):  \\
--- Proj.\ sep. ($^{\prime\prime}$) & --- & 1.058 $\pm$ 0.007 \\
--- Proj.\ sep. (AU) & --- & 55 $\pm$ 2\tablenotemark{j} \\
--- Position angle ($^\circ$) & --- & 56.0 $\pm$ 0.4 \\
\enddata
\tablecomments{Photometric values represent Subaru July 2012 measurements, unless noted otherwise.}
\tablenotetext{a}{Calculated using the published temperature from \citet{wu11} and evolutionary tracks from \citet{ekstr12}.}
\tablenotetext{b}{Calculated using the $H$-band magnitude, estimated \kapand{} age, and evolutionary models from \citet{chab00}.}
\tablenotetext{c}{\citet{fitz05}}
\tablenotetext{d}{\citet{wu11}}
\tablenotetext{e}{Based on measured colors and  \citet{cruz09} spectral identifications.}
\tablenotetext{f}{\citet{zuck11} and \citet{maro10}}
\tablenotetext{g}{Hipparcos; \citet{perr97}}
\tablenotetext{h}{2MASS; \citet{skru06}}
\tablenotetext{i}{Calculated by the authors, using 2MASS photometry and Hipparcos parallax.}
\tablenotetext{j}{Uncertainty is dominated by the host star parallax measurement.}
\end{deluxetable}

\end{document}